\documentstyle[sprocl]{article}
\def\Journal#1#2#3#4{{#1} {\bf #2}, #3 (#4)}

\def\be{\begin{equation}}
\def\ee{\end{equation}}
\def\bea{\begin{eqnarray}}
\def\eea{\end{eqnarray}}

\begin{document}
\title{NEUTRINO PHYSICS AND THE PRIMORDIAL ELEMENTAL
	ABUNDANCES}
\author{ CHRISTIAN Y. CARDALL AND GEORGE M. FULLER}
\address{Department of Physics, University of California, San Diego,
La Jolla,\\ CA 92093-0319, USA}
\maketitle\abstracts{
Limits can be placed on nonstandard neutrino physics
when big bang nucleosynthesis (BBN) calculations employing standard
neutrino physics agree with the
observationally inferred primordial abundances of deuterium (D), 
$^3$He, $^4$He,  and
$^7$Li.
These constraints depend most sensitively on the abundances of 
D and $^4$He.
New observational determinations of the primordial D and/or 
$^4$He abundances could force revisions in BBN constraints 
on nonstandard neutrino physics.
}

\section{Big Bang Nucleosynthesis and Neutrino Physics}
The primordial elemental abundances depend on $(n/p)_{\rm WFO}$, 
the ratio 
of neutrons to protons at ``weak freeze-out.'' Weak freeze-out occurs
when the expansion rate of the universe exceeds the rates 
of the reactions
that interconvert neutrons and protons. The $^4$He   
abundance depends 
strongly on $(n/p)_{\rm WFO}$, while the abundances of D, 
$^3$He, and $^7$Li  have a weaker
dependence on this quantity.

The expansion rate can be parametrized by an ``effective number
of neutrinos,'' $N_{\nu}$, which represents the relativistic
degrees of freedom in addition to those contributed by
photons and electron-positron pairs.
$N_{\nu}$ also indirectly parametrizes 
any phenomenon that affects the 
$^4$He abundance by
altering $(n/p)_{\rm WFO}$ 
(even though the effect may be on the $n \leftrightarrow p$
interconversion rates rather than through the expansion rate).

The primordial elemental abundances also depend on the 
baryon-to-photon
ratio $\eta$. The $^4$He yield depends somewhat
weakly on $\eta$, while the yields of the other 
light elements have
a strong dependence on $\eta$. The strength of the dependence of the
primordial elemental abundance yields on $N_{\nu}$  and $\eta$ 
can be visualized
from the figures in Refs. \cite{cardall1,fuller1}. 

Many investigators over the years have studied the 
effects of nonstandard
neutrino physics on BBN. These have included, for 
example, the number of
neutrinos; neutrino mixing with sterile neutrinos; 
adding a mass, lifetime,
and various decay products to the tau neutrino; and net cosmic lepton
number, i.e. endowing the neutrino seas with chemical potentials.

Standard big bang nucleosynthesis calculations assume three massless
neutrinos with zero net cosmic lepton number and no other relativistic
degrees of freedom. When a range of baryon-to-photon 
ratio $\eta$ is obtained
for which the calculated primordial elemental abundances 
agree with those
inferred from observations \cite{copi}, nonstandard neutrino physics 
can be constrained.
In the absence of a concordant range of $\eta$, one can 
either reasses the
observational errors or claim nonstandard neutrino physics 
(or some other physics) as a solution to bring the calculated 
abundaces into line with the 
primordial abundances inferred from observations. It should 
be noted that
nonstandard neutrino physics primarily affects 
$(n/p)_{\rm WFO}$, and could therefore
change $^4$He rather easily. But more extreme neutrino physics
would have to be introduced to significantly 
adjust the D, 
$^3$He, and $^7$Li abundances,
and then fine tuning would be  
required to avoid unacceptable changes in the calculated
$^4$He abundance yield \cite{cardall1}.

\section{The Observational Situation}
The observationally inferred abundances 
of the light elements produced in BBN have continued to be the subject
of discussion and observational effort. Of particular interest are
deuterium (D), which provides the most sensitive measure of the baryon
density; and $^4$He, which constrains $N_{\nu}$. Also of interest is
the primordial abundance of $^7$Li, which provides a 
cross-check to the
range of $\eta$ determined by D and $^4$He. The primordial abundance
of $^3$He is less useful for the purpose of determining 
$\eta$ due to
its complicated history of chemical and galactic evolution.

While inferences of the primordial D/H ratio formerly were based on 
``local'' measurements in the solar system and interstellar medium,
observations of quasar absorbtion systems (QAS) have recently provided
determinations of D/H in a much more pristine environment. Initially
there was a strong dichotomy in inferences of D/H, with claimed values
differing by an order of magnitude \cite{cardall1}. However, at this
conference the situation appears to be coming to a resolution, with
the low value of D/H now favored \cite{hogan1}.
It is also worth pointing out that 
the higher baryon density
implied by low D/H is advantageous from other 
cosmological/astrophysical
considerations \cite{cardall1,fuller1}.

The inferred primordial abundance of $^4$He has also 
been the subject of
recent action. A commonly accepted range of the $^4$He 
abundance\cite{olive1} does
not provide a value of $\eta$ concordant with that inferred from the
low values of D/H inferred from QAS \cite{cardall1}, for $N_{\nu}$=3. 
(As it turns out, 
the primordial abundance of $^7$Li inferred from the 
``Spite plateau'' does
not agree with low D/H for $N_{\nu}$=3 either.)
One group of 
observers has claimed \cite{izotov}  
a higher range of $^4$He abundance, but this analysis may have some 
difficulties \cite{olive1}. It is not unreasonable, however, 
that systematic errors in the 
determinations of $^4$He and $^7$Li could
allow for concordance at low D/H for $N_{\nu}$=3.

\section{An Example: Mixing with Sterile Neutrinos}
Neutrino mixing with ``sterile'' (SU(2) singlet) neutrinos
during the epoch of BBN, which has been studied by many
authors \cite{cardall2}, 
provides an example of how constraints on nonstandard 
neutrino physics depend on the primordial abundances of D
and $^4$He. Two kinds of mixing with sterile neutrinos have
been suggested to solve neutrino puzzles: $\nu_e \leftrightarrow
\nu_s$ for the solar neutrino problem, 
and $\nu_{\mu} \leftrightarrow
\nu_s$ for the atmospheric neutrino problem.

Neutrino mixing with steriles does two things to increase primordial
$^4$He: first, it effectively brings another degree of freedom into 
thermal contact; and second, it depletes the $\nu_e$ population,
reducing the $n \leftrightarrow p$ interconversion rates. The demand
that $^4$He not be overproduced places limits on mixing with steriles.
The constraint comes from upper bounds on D/H and the $^4$He mass
fraction. This is because 
the largest D/H (and hence the minimum allowed
$\eta$) allows the largest $N_{\nu}$ for a given $^4$He abundance.

In constraining mixing with steriles, previous studies assumed, for
example, $\eta > 2.8 \times 10^{-10}$ based on estimates of primordial
(D+$^3$He)/H derived from ``local'' measurements. Until 
this conference,
it appeared that discordant deuterium determinations could be 
indicating a minimum value of $\eta$ that was either lower or
higher than this. As an example of the consequences of a possible
change in the minimum value of $\eta$, Cardall and Fuller 
published a calculation \cite{cardall2} 
in which the $\nu_{\mu} \leftrightarrow \nu_s$ solution 
to the atmospheric
neutrino problem was {\em assumed}, and the resultant $^4$He 
mass fraction
was plotted as a function of D/H. For the purported ``high'' values of
D/H, it appeared that the $\nu_{\mu} \leftrightarrow \nu_s$ 
solution to the atmospheric neutrino problem might be allowed.
For low D/H, this mixing is even more strongly ruled out.

\end{document}